\documentclass[10pt]{article}
\input epsf

\begin{document}

\title{Note on the Sunyaev-Zel'dovich thermal effect }
\author{{\small A. Sandoval-Villalbazo}$^{a}${\small \ and L.S. Garc\'{\i}a-Col\'{\i}n}
$^{b,\,c}$ \\
$^{a}${\small \ Departamento de F\'{\i}sica y Matem\'{a}ticas, Universidad
Iberoamericana }\\
{\small \ Lomas de Santa Fe 01210 M\'{e}xico D.F., M\'{e}xico }\\
{\small \ E-Mail: alfredo.sandoval@uia.mx }\\
{\small \ }$^{b}${\small \ Departamento de F\'{\i}sica,
Universidad Aut\'{o}noma Metropolitana }\\
{\small \ M\'{e}xico D.F., 09340 M\'{e}xico }\\
{\small \ }$^{c}${\small \ El Colegio Nacional, Centro
Hist\'{o}rico 06020 }
\\
{\small \ M\'{e}xico D.F., M\'{e}xico }\\
{\small \ E-Mail: lgcs@xanum.uam.mx}}
\maketitle

\begin{abstract}
{\small In a previous publication we have derived an expression
for the full distorted spectrum arising when the photons of the
cosmic background radiation are absorbed and emitted by an
optically thin gas. The expression simply adds up the effects of
the joint probability that an electron absorbs an incoming photon
of frequency }$\bar{\nu}${\small \ and emits it with a frequency
}$\nu ${\small . We here show that such an expression is identical
to the Kompaneets equation describing the thermal
Sunyaev-Zel'dovich effect. This clearly shows that the two
pictures, diffusion and absorption-emission are mathematically
equivalent to first order in the Compton parameter. Other
consequences of this result are also discussed.}
\end{abstract}

In a recent publication \cite{one} we have given a detailed
discussion of how the thermal Sunyaev-Zel'dovich (SZ) effect
\cite{two} \cite{three} may be interpreted in terms of a rather
simplified picture as an absorption-emission process. In fact, our
view is that such an effect is indeed that of an
absorption-emission process in which a few photons happen to be
captured by electrons in the optically thin gas. An electron
moving with a given thermal velocity emits (scatters) a photon
with a certain incoming frequency $\bar{ \nu}$ and outgoing
frequency $\nu $. The line breath of this process is readily
calculated from kinetic theory taking into account that the media
in which the process takes place has a small optical depth
directly related to the Compton parameter $y$. When the resulting
expression, which will be called \emph{structure factor}, is
convoluted with the incoming flux of photons obtained from
Planck's distribution, one easily obtains the disturbed spectra.

In Ref. \cite{one} we derived an analytical expression for the full
distorted spectrum of the scattered radiation which reads as
\begin{equation}
I\left( \nu \right) =\frac{1}{\sqrt{\pi }W\left( \nu \right) }
\int_{0}^{\infty }I_{o}\left( \bar{\nu}\right) \exp \left[ -\left(
\frac{ \bar{\nu}-(1-2y)\,\nu }{W\left( \nu \right) }\right)
^{2}\right] \,d\bar{\nu} \label{uno}
\end{equation}
and defines the joint probability of finding an electron scattering a photon
with incoming frequency $\bar{\nu}$\thinspace and outgoing frequency $\nu $
multiplied by the incoming photon flux $I_{o}\left( \bar{\nu}\right) $. In
Eq. (\ref{uno}) the incoming flux is given by
\begin{equation}
I_{o}\left( \bar{\nu}\right)
=\frac{2h\bar{\nu}^{3}}{c^{2}}\frac{1}{e^{\frac{h
\bar{\nu}}{kT}}-1}  \label{dos}
\end{equation}
where $T\equiv 2.726\,\,K$, the temperature of the cosmic
background radiation, $k$ is Boltzmann's constant, $h$ is Planck's
constant and $c$ the speed of light. Further, $W\left( \nu \right)
$ is the width of the spectral line at frequency $\nu $  and its
squared value reads (Eq. (14) in Ref. [1]):
\begin{equation}
W^{2}\left( \nu \right) =\frac{4kT_{e}}{m_{e}c^{2}}\tau \,\nu
^{2}=4\,y\,\nu^{2}  \label{tres}
\end{equation}
where $T_{e}$ is the temperature of the electron scatterers, $m_{e}$ is the
electron mass and $\tau $ is the optical depth, whose meaning is fully
discussed in \cite{one}.

The purpose of this note is simply to show that Eq. (\ref{uno})
can be evaluated analytically in an almost trivial way leading to
an expression which is identical to the one reported in the
literature. Setting in Eq. (\ref{uno})
\begin{equation}
\alpha =\frac{\bar{\nu}-(1-2y)\,\nu }{W\left( \nu \right) }  \label{cuatro}
\end{equation}
we get that
\begin{equation}
I\left( \nu \right) =\frac{1}{\sqrt{\pi }}\int_{-\infty }^{\infty
}I_{o}\left( \nu+\Delta \nu \right) \exp \left[ -\alpha
^{2}\right] \,d\alpha  \label{cinco}
\end{equation}
where
\begin{equation}
\Delta \nu =-2\,y\,\nu+2y^{1/2}\nu \alpha   \label{seis}
\end{equation}
Notice should be made on the fact that the lower limit in the
integral can be readily estimated to be a large negative number,
which is \ for all practical purposes $-\infty $. The trick is now
immediate. Since $y$ is a very small number \cite{four} and
$I_{o}$ is an analytical function of $\nu $ we may expand the
integrand in Eq. (\ref{cinco}) in a Taylor series. Since $
\Delta\nu $ is linear in $ \alpha $, all odd powers of $ \alpha $
yield zero upon integration so that after a straightforward
calculation we get that
\begin{equation}
I\left( \nu \right) -I_{o}\left( \nu\right) =-2y\,\nu
\frac{\partial I_{o}}{
\partial \nu }+y\,\nu ^{2}\frac{\partial ^{2}I_{o}}{\partial \nu ^{2}}
+2y\,^{2}\nu ^{2}\frac{\partial ^{2}I_{o}}{\partial \nu ^{2}}-2y\,^{2}\nu
^{3}\frac{\partial ^{3}I_{o}}{\partial \nu ^{3}}+...  \label{ocho}
\end{equation}
To first order in $y$ \ we reproduce \emph{exactly} the analytic
expression for the distortion curve \cite{two}-\cite{three}. It is
also interesting to notice that, taking into  account the second
order terms in $y$ included in Eq. (\ref{ocho}), the crossover
frequency $\nu _{c}$ corresponding to the solution of equation
\begin{equation}
I\left( \nu \right) -I_{o}\left( \nu\right) =0  \label{nueve}
\end{equation}
is now clearly dependent on the value of $\,y$ \ namely, on the
electron temperature, a feature not ordinarily recognized for a
non-relativistic intra-cluster gas. For a typical value of
$\,y=10^{-4}$, we obtain $\nu _{c}=217.16$ GHz, while the value
obtained by considering only linear terms in $y$ yields $ \nu
_{c}=217.12$ GHz, a difference of nearly $40$ MHz.

The result expressed in Eq. (\ref{ocho}) is of significant
importance. Not only shows that laborious calculations may be
expressed in terms of rather elementary concepts borrowed from
statistical mechanics, but clarifies an until now intriguing
result. Indeed, Eq. (\ref{ocho}) contains the famous Kompaneets
equation \cite{five}-\cite{seven} describing the thermal SZ effect
using a diffusive mechanism to account for the motion of photons
in an optically thin electron gas where the electron temperature
$T_{e}>>T_{CMB}$, the photon temperature (see specifically Ref.
\cite{six}). One must notice that, since most photons \emph{are
not scattered even once}, a diffusion approximation would then
hardly seem adequate \cite{four}. Nevertheless, the diffusion
mechanism is mathematically identical to the absortion-emission
process of photons by electrons to first order in $y$. The problem
posed in Ref. \cite{one} requiring an explanation of why two
different mechanisms lead to the same result is now mathematically
solved. The exponential function in the integrand of
Eq.(\ref{uno}) is playing the role of a Green's function for the
diffusion equation. Finally, we want to stress that it is now
clear why the curves describing the thermal SZ effect obtained by
numerical integration of different mathematical expressions turns
to be identical. We therefore expect that the relativistic SZ
effect and other photon scattering problems in hot plasmas can be
handled by a similar procedure.

The authors wish to thank G.H. Weiss for kindly supplying the idea behind
the integration method. This work has been supported by CONACyT (Mexico),
project 41081-F.


\begin{thebibliography}{1}
\bibitem[1]{one} A. Sandoval-Villalbazo and L.S. Garc\'{\i}a-Col\'{\i}n;
J.Phys. A: Math and Gen. \textbf{36 }(2003) 4641-4650   [astro-ph/0208440].

\bibitem[2]{two} Y.B. Zel'dovich and R.A. Sunyaev, Astrophys. Space Sci.
\textbf{4}, 301 (1969).

\bibitem[3]{three} R.A. Sunyaev and Ya. B. Zel'dovich. Comm. Astrophys.
Space Phys. \textbf{4}, 173 (1972).

\bibitem[4]{four} Y. Rephaeli, Astrophys. J. \textbf{445}, 33 (1995);\newline
A.D. Dolgov, S.H. Hansen, D.V. Semikoz and S. Pastor, Astrophys.
J. \textbf{ 554} 74 (2001) \ [astro-ph/0010412].

\bibitem[5]{five} P.J.E. Peebles, Physical Cosmology (Princeton Univ. Press,
Princeton, 2nd. Ed. 1993).

\bibitem[6]{six} R.A. Sunyaev and Ya. B. Zel'dovich; Ann. Rev. Astrom.
Astrophys. \textbf{18}, 537 (1980)

\bibitem[7]{seven} A.S. Kompaneets; JETP. \textbf{4}, 730 (1957).
\end{thebibliography}
\end{document}